\documentclass[11pt]{article}

\usepackage[preprint]{acl}

\usepackage{times}
\usepackage{latexsym}

\usepackage[T1]{fontenc}

\usepackage[utf8]{inputenc}

\usepackage{microtype}

\usepackage{inconsolata}

\usepackage{graphicx}

\usepackage{tabularx}
\usepackage{booktabs}
\usepackage{float}
\usepackage[style=american]{csquotes}
\usepackage{amsmath, amssymb}

\newcommand{\updownarrows}{\mathbin\uparrow\hspace{-.5em}\downarrow}

\title{Vaporizer: Breaking Watermarking Schemes for Large Language Model Outputs}

\author{Jonathan Hong Jin Ng \and Anh Tu Ngo \and Anupam Chattopadhyay \\
        College of Computing and Data Science \\ Nanyang Technological University \\ Singapore}

\begin{document}
\maketitle
\begin{abstract}
In this paper, we investigate the recent state-of-the-art schemes for watermarking large language models (LLMs) outputs. These techniques are claimed to be robust, scalable and production-grade, aimed at promoting responsible usage of LLMs. We analyse the effectiveness of these watermarking techniques against an extensive collection of modified text attacks, which perform targeted semantic changes without altering the general meaning of the text content. Our approach encompasses multiple attack strategies, which include lexical alterations, machine translation, and even neural paraphrasing. The attack efficacy is measured with two target criteria - successful removal of the watermark and preservation of semantic content. We evaluate semantic preservation through BERT scores, text complexity measures, grammatical errors, and Flesch Reading Ease indices. The experimental results reveal varying levels of effectiveness among different watermarking models, with the same underlying result that it is possible to remove the watermark with reasonable effort. This study sheds light on the strengths and weaknesses of existing LLM watermarking systems, suggesting how they should be constructed to improve security of available schemes.
\end{abstract}

\section{Introduction}
The emergence of large language models (LLMs) has transformed text generation capabilities in industries such as automated assistants, content creation or even code generation \cite{IBM2023LLMs}. However, the adoption of these models has led to concerns regarding content authenticity, plagiarism detection, and model attribution \cite{GoogleCloud2023Watermarks}. As a result, watermarking techniques have been suggested to embed invisible marks within the generated text to verify authorship \cite{Liu2024Survey}. Unfortunately, efforts to verify and detect watermark signals remain vulnerable to semantic text manipulations that manipulate text but preserves semantic integrity, effectively destroying the watermarks \cite{Hou2024SemStamp}. Understanding these vulnerabilities is crucial for reinforcing the reliability of the watermarking methods in real-world applications.

A large language model (LLM) is a neural network trained on vast textual data to learn a probability distribution over sequences of words. Formally, given a sequence of tokens \(X=(x_1, x_2, ...x_n)\), an LLM models the conditional probability distribution
\(P(X) = \prod_{i=1}^{n} P(x_i | x_1, x_2, ..., x_{i-1})\), which represents the likelihood of token \(x_i\) occurring given its previous tokens. These models utilise deep architectures, such as transformer~\cite{vaswaniAttentionAllYou2017}, to capture long-term dependencies and produce coherent text. 

Watermarks change the probability distribution of the text, creating recognizable patterns that adversarial approaches seek to detect and disrupt. These schemes have evolved from naive lexical modifications to sophisticated probabilistic embedding techniques, aiming to ensure reliable detection while minimising the impact on text quality, with various schemes employing different embedding methods as shown in Table~\ref{table:llm_watermarking_evolution_types}. The selection of the three watermarking schemes for evaluation stems from their state-of-the-art status and their claims of robustness, scalability, and practical viability in watermarking LLM outputs. The chosen schemes Provable Robust Watermarking~\cite{zhao2024provable}, Publicly Detectable Watermarking~\cite{fairoze2025publicly}, and SynthID~\cite{deepmind2024scalable} represent distinct approaches: one emphasising probabilistic distribution control, another leveraging cryptographic signatures, and the last using tournament-based selection. Evaluating all possible watermarking methods is not possible, and thus these three serve as strong representatives of current advancements, providing a meaningful analysis of vulnerabilities without redundancy. 

\begin{table}[h]
\centering
\begin{tabularx}{0.5\textwidth}{Xr}
\toprule
\textbf{Watermarking Scheme} & \textbf{Type} \\
\midrule
SynthID~\cite{deepmind2024scalable} & Statistical\\
Publicly Detectable~\cite{fairoze2025publicly} & Cryptographic\\
Unigram Watermarking~\cite{zhao2024provable} &  Statistical\\
Undetectable Watermark~\cite{cryptoeprint:2023/763} &  Cryptographic\\
Token Partitioning~\cite{kirchenbauer2023watermark} & Statistical\\
\bottomrule
\end{tabularx}
\caption{Evolution of LLM Watermarking Schemes with Types (Unexhaustive)}
\label{table:llm_watermarking_evolution_types}
\end{table}

This work provides the first comprehensive demonstration that leading watermarking schemes for LLM outputs are vulnerable to a wide range of semantic-preserving attacks. We empirically show that watermark signals—whether statistical or cryptographic—can be removed with high success rates through methods that maintain the semantic integrity of the original content. In particular, our evaluation reveals that neural paraphrasing techniques are especially effective at degrading watermark detectability while preserving linguistic quality. While our primary contribution lies in empirically proving the fragility of current watermarking schemes, several secondary contributions naturally follow from our study:

\begin{itemize}
\item We develop a systematic evaluation framework encompassing multiple attack vectors and detection metrics.

\item We quantify the trade-off between watermark removal effectiveness and text quality degradation.

\item We analyze the impact of incremental text modifications on both watermark detection and semantic preservation.

\item We provide insights and recommendations for designing more resilient watermarking systems.
\end{itemize}

These contributions collectively advance our understanding of watermarking robustness in real-world scenarios and provide valuable guidance for developing more secure content authentication technologies for AI-generated text.

\section{Related Work}
This section provides a comprehensive analysis of the three chosen watermarking schemes, examining their theoretical foundations and their implementation mechanisms. This review establishes the necessary foundation for understanding the vulnerability assessment presented in subsequent sections.

Kirchenbauer et al.~\cite{kirchenbauer2023watermark} proposed the idea of using sets of ``green'' and ``red'' tokens. During the sampling process, the output logits are manipulated so that the green tokens are sampled more. This paper discusses a way to adaptively handle the output tokens in both low-entropy and high-entropy contexts. Zhao et al.~\cite{zhao2024provable} introduced a statistical bias during text generation by partitioning the model's vocabulary into two disjoint subsets: a "green list" of favored tokens and a "red list" of disfavored tokens. During generation, the model is encouraged to select tokens from the green list more frequently, thereby embedding a statistical watermark without perceptibly degrading fluency or coherence. To detect the watermark, one analyzes the proportion of green list tokens in the generated text. A z-score is computed to quantify the deviation of this proportion from the expected baseline under normal, unwatermarked generation. If the z-score exceeds a predetermined threshold (typically near zero), the text is considered to contain a watermark. This scheme is robust to minor perturbations but remains susceptible to more sophisticated semantic-preserving transformations that redistribute token frequencies.


Work from Fairoze et al.~\cite{fairoze2025publicly} employs a cryptographic approach that allows any verifier to detect a watermark without requiring access to the original model or secret keys. The process begins by generating a segment of text and encoding a digital signature within it using hash functions and error-correcting codes. Specific tokens are selectively sampled such that the resulting sequence encodes a verifiable watermark. Detection is conducted by reconstructing the original hash values and attempting to recover the encoded signature from the text. If a valid signature is successfully recovered and verified against public parameters, the text is declared watermarked. While the method offers transparency and public verifiability, its dependence on precise token-level structure makes it particularly vulnerable to even minor semantic-preserving alterations such as paraphrasing or translation.


SynthID~\cite{deepmind2024scalable}, developed by Google DeepMind, uses a hybrid watermarking strategy based on tournament sampling. For each token to be generated, the language model samples multiple candidate tokens. A deterministic scoring function, seeded by a secret key, ranks these candidates based on computed values derived from their context and identity. The highest-ranking candidate is selected, embedding a watermark through subtle but consistent preferences. Detection proceeds by re-evaluating the generated sequence using the same scoring function and computing the average score across all tokens. If this score exceeds a defined threshold (typically around 0.5), the sequence is classified as watermarked. SynthID offers a balance between subtlety and robustness, but its reliance on repeated structural patterns makes it vulnerable to aggressive paraphrasing and token-level shuffling.

\section{Methodology}

This section details our systematic approach to evaluating watermarking robustness against text manipulations. We first present our multi-faceted attack framework, which encompasses lexical transformations, machine translation techniques, and neural paraphrasing models. We then describe our comprehensive evaluation metrics that assess both watermark removal efficacy and text quality preservation. Finally, we outline our experimental setup, including dataset selection, implementation details, and parameter configurations.



\begin{figure*}[htbp]
\centering
\includegraphics[width=0.9\linewidth]{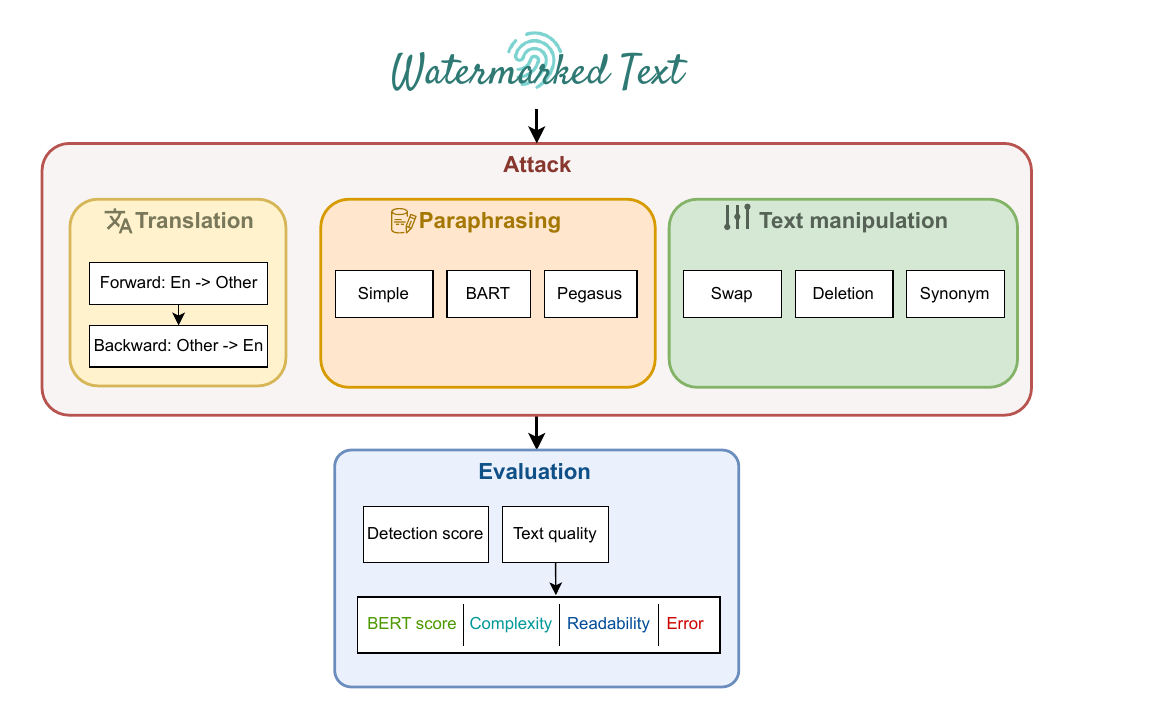}
\caption{Hierarchical structure of attack strategies implemented in our framework. The top right represents single attack flow while the bottom right represents the incremental attack flow. This pipeline enables systematic, incremental modifications to watermarked text while capturing metrics on both watermark detectability and semantic preservation.}
\label{fig:attack_framework}
\end{figure*}

Our attack methodology (Figure~\ref{fig:attack_framework}) encompasses three distinct categories of watermark removal strategies, each implemented to balance effectiveness and text quality preservation. These attacks are performed on the watermarked text (For Publicly Detectable Watermarking scheme, the initial segment character bits used to generate the signature is preserved and the attacks are performed on the remaining character bits).

The lexical transformation module implements four distinct approaches to text modification. The first is synonym replacement, which uses WordNet's semantic database to substitute words with synonyms that maintain the original meaning. Our word deletion attack uses an advanced random sampling method for token removal from the given text. Each deletion operation is tracked, allowing us to analyse the relationship between specific word removals and watermark detection scores. Additionally, we perform random swapping of words to randomly switch positions of any two given tokens. The algorithm for the adjacent word swap attack makes controlled local perturbations to the order of words. The system logs every substitution and the resulting change in watermark detection and coherence, thus informing us of how word order contributes to watermark persistence. Other than implementing one-off attack vectors, incremental attacks are also performed on the watermarked text for these lexical transformations to understand the relationship between incremental attacks and watermark quality as well as text quality preservation.

Our translation attack system utilises the MarianMT neural machine translation framework, implementing three primary language pairs: English-French, English-German, and English-Spanish. Each translation operation is performed bidirectionally, first translating the text to the target language and then back to English. 

The framework implements two approaches to neural paraphrasing. The first is based on the BART model (facebook/bart-large-cnn), which applies beam search with a beam size of 4 and a temperature setting of 0.7. The second approach uses the Pegasus paraphrase model, which has different structural traits and ways of producing outputs.

Our evaluation system combines rigorous watermark detection analysis with comprehensive text quality assessment. The watermark detection component tracks the detection scores for both original and attacked texts, generating detailed statistics about the success rates of the attack and the degradation patterns of the watermarks. The system also analyzes how the detectability of watermarks changes with incremental text modifications. The text quality analysis system employs multiple complementary approaches to assess semantic preservation and text quality. At its core, it uses the DeBERTa-XLarge-MNLI model to compute BERT F1 scores. The system also calculates readability metric Flesch Reading Ease scores to assess the impact of attacks on text comprehensibility. Grammar analysis is performed using LanguageTool, providing detailed error detection and classification. For translation-based attacks, additional quality metrics assess adequacy and fluency, combining BERT scores with ROUGE metrics to provide a complete picture of translation quality. Additionally, the framework generates comprehensive visualisations and reports, including violin plots of detection score distributions, quality metric comparisons, and detailed analysis of the trade-offs between attack effectiveness and text quality preservation.

The experimental approach was built on top of a complete infrastructure written in Python and makes use of the PyTorch library for GPU acceleration via CUDA. All experiments were run on an NVIDIA RTX 4070 Super GPU with 12GB of VRAM, which adequately supported the computational demands for the watermarking and attack processes in terms of GPU resources. The experiments were done using an isolated system with Python version 3.11. We utilised a dataset comprising 100 diverse text prompts generated using ChatGPT, using the following prompt:

\begin{displayquote}
\textit{``Create a comprehensive list of 100 general writing prompts suitable for students or writers. Include a variety of prompt types such as creative writing tasks, explanations of concepts, descriptive exercises, and opinion pieces. Format the output as a JSON object with numbered keys from 1 to 100, with each value being a complete writing prompt. Include prompts about nature, technology, society, personal experiences, education, health, environment, and other diverse topics.''}
\end{displayquote}

These prompts encompass a wide range of text styles, complexities and domains, ensuring that the evaluation captured the performance of the watermarking scheme across different contexts. Each prompt was executed on the respective watermarking systems to yield watermarked outputs, which acted as the basis for our attack evaluations. In addition, to create a control group, we obtained unwatermarked outputs from each language model using the same model framework but with disabling watermarking controls. This strategy allowed us to compute baseline detection scores and quality metrics.

To execute our attack, we set each transformation approach to parameters that we believed achieved an optimal balance between attack efficacy and text quality preservation. In the case of lexical transformations, synonym replacement processes were restricted to content words (nouns, verbs, adjectives, and adverbs) with a maximum substitution rate of 30\% to preserve coherence. In deletion-based word attacks, we employed a controlled random sampling strategy where up to 10\% of tokens were removed, focusing on function words to lessen semantic impact. Adjacent word swap operations were limited to shifts within pairs who share the same syntactic category to avoid breaking grammatical rules. In translation-based attacks, we implemented pre-trained MarianMT models from the Hugging Face Transformers library, setting batch and beam sizes at 8 and 4, respectively, for both translation directions. BART (facebook/bart-large-cnn) and Pegasus models were used for neural paraphrase attack with BART set to a temperature of 0.7 which provided the right amount of diversity and cohesiveness to the paraphrases. Controlled with a maximum sequence length of 512 tokens, both systems processed longer texts by segmenting them into overlapping chunks.

\section{Results and analysis}
This section presents our empirical findings on the vulnerability of watermarking schemes against various attack strategies. We analyze each watermarking approach separately, examining attack effectiveness, text quality impact, and the relationship between transformation intensity and watermark degradation. Through statistical analysis and visualisations, we identify key patterns in watermark robustness and text quality trade-offs across different attack vectors.

Figure~\ref{fig:watermark_analysisa} display text quality metrics (BERT scores, complexity, readability, and grammatical errors) for each attack strategy. Figure~\ref{fig:watermark_analysisb} shows the Z-score distributions for watermarked text under different attacks, where lower values indicate more successful watermark removal—notably, Pegasus Paraphrase achieved the lowest median Z-score (1.21) with 23\% of samples falling below the detection threshold. Figure~\ref{fig:watermark_analysisc} demonstrate the incremental effects of increasing attack intensity on detection scores, revealing how watermark detectability progressively degrades as modification percentage increases. These visualisations confirm that neural paraphrasing techniques achieve the best balance between watermark removal effectiveness and text quality preservation for Provable Robust Watermarking.

\begin{table*}[ht]
\centering

\label{tab:unigram_comprehensive_analysis}
\begin{tabularx}{\textwidth}{Xcccccc}
\toprule
\textbf{Attack Method} & \textbf{\% Success Rate} & \textbf{Mean Z-Score} & \textbf{Std Dev} & \textbf{Min} & \textbf{Max} & \textbf{Median} \\
\midrule
Watermarked (Ref.) & --- & 8.43 & 2.97 & 0.24 & 13.77 & 8.62 \\
Unwatermarked & --- & -0.52 & 1.39 & -6.93 & 2.51 & -0.40 \\
Transl. (German) & 5.0 & 6.30 & 3.48 & -1.90 & 19.67 & 6.29 \\
Transl. (Spanish) & 5.0 & 6.46 & 4.15 & -1.00 & 22.45 & 6.33 \\
Transl. (French) & 4.0 & 6.03 & 3.70 & -1.14 & 22.45 & 6.20 \\
Simple Paraphrase & 4.0 & 3.52 & 2.02 & -1.00 & 9.81 & 3.64 \\
BART Paraphrase & 2.0 & 4.22 & 1.67 & -2.89 & 6.53 & 4.58 \\
Pegasus Paraphrase & 23.0 & 0.93 & 1.14 & -1.67 & 3.54 & 1.21 \\
Swap & 4.0 & 7.31 & 3.10 & -0.71 & 13.49 & 7.81 \\
Deletion & 3.0 & 6.08 & 2.67 & -1.00 & 11.17 & 6.39 \\
Syn. Replacement & 1.0 & 5.46 & 2.77 & -0.71 & 10.95 & 6.05 \\
\bottomrule
\end{tabularx}
\caption{Watermark Preservation Analysis of Provable Robust Watermarking Watermarking scheme. Note that "---" is for Watermarked and Unwatermarked as those are not attacks.}
\label{table:watermark_preservation}
\end{table*}

Table~\ref{table:watermark_preservation} quantifies the effectiveness of various attack strategies applied to Provable Robust Watermarking watermarked text, where success rates indicate the percentage of attacked samples with Z-scores below 0. The distribution of the mean z-score and associated statistics reveal the distribution characteristics of detection scores across the samples. For Provable Robust Watermarking scheme, Pegasus Paraphrase achieved the highest success rate (23\%) with the lowest mean Z-score (0.93), significantly outperforming other attack methods. Translation-based approaches showed moderate effectiveness (4-5\% success rate), while lexical transformations like synonym replacement proved least effective (1-4\% success rate). For comparison, watermarked text had a mean Z-score of 8.43, while unwatermarked text averaged -0.52.

\begin{table*}[h]
\centering
\label{tab:comprehensive_impact}
\begin{tabularx}{\textwidth}{Xcccccccc}
\toprule
\textbf{Attack Method} & \multicolumn{2}{c}{\textbf{BERT Score}} & \multicolumn{2}{c}{\textbf{Complexity}} & \multicolumn{2}{c}{\textbf{Readability}} & \multicolumn{2}{c}{\textbf{Error}} \\
\cmidrule(lr){2-3}\cmidrule(lr){4-5}\cmidrule(lr){6-7}\cmidrule(lr){8-9}
 & \textbf{Avg} & \textbf{\% $\updownarrows$} & \textbf{Avg} & \textbf{\% $\updownarrows$} & \textbf{Avg} & \textbf{\% $\updownarrows$} & \textbf{Avg} & \textbf{\% $\updownarrows$} \\
\midrule
Watermarked (Ref.) & 1.00 & --- & 5.95 & --- & 78.89 & --- & 3.44 & --- \\
Unwatermarked & --- & --- & 6.34 & 6.55 & 79.47 & 0.73 & 1.40 & -59.30 \\
Transl. (German) & 0.96 & -4.27 & 6.56 & 10.25 & 78.10 & -1.00 & 2.31 & -32.85 \\
Transl. (Spanish) & 0.96 & -4.20 & 7.18 & 20.67 & 73.47 & -6.87 & 3.79 & 10.17 \\
Transl. (French) & 0.95 & -5.17 & 7.32 & 23.03 & 67.84 & -14.00 & 2.28 & -33.72 \\
Simple Paraphrase & 0.83 & -16.79 & 9.24 & 55.29 & 63.42 & -19.61 & 37.99 & 1004.36 \\
BART Paraphrase & 0.92 & -7.78 & 5.98 & 0.50 & 81.45 & 3.25 & 1.05 & -69.48 \\
Pegasus Paraphrase & 0.87 & -12.61 & 5.46 & -8.24 & 83.96 & 6.43 & 0.03 & -99.13 \\
Swap & 0.92 & -8.15 & 6.17 & 3.70 & 79.20 & 0.40 & 9.47 & 175.29 \\
Deletion & 0.91 & -9.10 & 5.54 & -6.89 & 79.13 & 0.31 & 5.23 & 52.03 \\
Syn. Replacement & 0.90 & -10.08 & 7.81 & 31.26 & 69.86 & -11.45 & 30.88 & 797.67 \\
\bottomrule
\end{tabularx}
\caption{Comprehensive Text Quality Analysis by Attack Method of Provable Robust Watermarking Scheme. The point of reference for the \% change is the original watermarked texts.}
\label{table:text_quality}
\end{table*}

Table~\ref{table:text_quality} presents how text quality metrics are affected for each attack method applied to Provable Robust watermarked text. Mean BERT Score quantifies semantic preservation, with percentage change relative to the original watermarked text. Mean Complexity (Gunning Fog Index) measures linguistic complexity, while Mean Readability (Flesch Reading Ease) assesses comprehensibility, and Mean Errors reports the average grammatical error count per sample. Translation-based approaches best preserved semantic content (4-5\% BERT reduction) but increased text complexity (10-23\%). Neural paraphrasing methods (Bart, Pegasus) showed moderate semantic degradation (7-12\% BERT reduction) while maintaining reasonable error rates. Simple paraphrase and synonym replacement caused the most severe quality degradation, with BERT reductions of 10-16\% and grammatical error increases of 797-1004\%, highlighting the critical trade-off between watermark removal effectiveness and text quality.

Figures~\ref{fig:text_qualitya}, ~\ref{fig:text_qualityb}, ~\ref{fig:text_qualityc}, ~\ref{fig:text_qualityd} show the incremental effects of text modifications on Provable Robust Watermarking text quality metrics of four attack types: word swapping, word deletion, synonym replacement, and simple paraphrasing respectively. Each figure shows a different quality dimension: (1) text complexity, (2) BERT scores measuring semantic preservation, (3) readability, and (4) grammatical error counts. The error bands represent 95\% confidence intervals across 100 test samples. Results reveal a clear text quality degradation, with BERT scores (second row) showing near-linear degradation as modification intensity increases. Notably, synonym replacement exhibits the steepest quality degradation curve, while deletion maintains better semantic coherence.

Figure~\ref{fig:watermark_analysis4} showcases the comprehensive analysis of SynthID Watermarking vulnerability across attack methods. Figure~\ref{fig:watermark_analysis5} illustrates g-score distributions across attack methods, with lower g-scores indicating more effective watermark removal. Figure~\ref{fig:watermark_analysis6} track the incremental effects of increasing attack intensity on detection scores. These visualisations collectively demonstrate that SynthID requires more aggressive text transformations for successful watermark removal than Provable Robust Watermarking, with Pegasus Paraphrase offering the most favorable balance between removal effectiveness.

\begin{table*}[h]

\centering
\begin{tabularx}{\textwidth}{Xccccccc}
\toprule
\textbf{Attack Method} & \textbf{\% Success Rate} & \textbf{Mean Score} & \textbf{Std Dev} & \textbf{Min} & \textbf{Max} & \textbf{Median} \\
\midrule
Watermarked (Ref.) & --- & 0.563 & 0.017 & 0.510 & 0.600 & 0.564 \\
Unwatermarked & --- & 0.501 & 0.009 & 0.474 & 0.536 & 0.500 \\
Transl. (German) & 2.0 & 0.526 & 0.013 & 0.477 & 0.551 & 0.527 \\
Transl. (Spanish) & 2.0 & 0.530 & 0.016 & 0.443 & 0.562 & 0.530 \\
Transl. (French) & 3.0 & 0.527 & 0.017 & 0.477 & 0.594 & 0.528 \\
Simple Paraphrase & 5.0 & 0.512 & 0.008 & 0.498 & 0.543 & 0.510 \\
BART Paraphrase & 3.0 & 0.544 & 0.023 & 0.488 & 0.594 & 0.544 \\
Pegasus Paraphrase & 14.0 & 0.538 & 0.039 & 0.425 & 0.650 & 0.541 \\
Swap & 9.0 & 0.512 & 0.009 & 0.492 & 0.538 & 0.512 \\
Deletion & 3.0 & 0.514 & 0.009 & 0.477 & 0.540 & 0.513 \\
Syn. Replacement & 11.0 & 0.510 & 0.009 & 0.490 & 0.540 & 0.510 \\
\bottomrule
\end{tabularx}
\caption{Comprehensive Analysis of SynthID Watermarking Attack Methods}
\label{table:synthid_attack}
\end{table*}

Table~\ref{table:synthid_attack} quantifies the effectiveness of various attack strategies applied to SynthID watermarked text, where success rates indicate the percentage of attacked samples with g-scores below 0.5. The distribution of the mean g-score and associated statistics reveal the distribution characteristics of detection scores across the samples. For SynthID Watermarking scheme, Pegasus Paraphrase likewise achieved the highest success rate (14\%), significantly outperforming other attack methods. Translation-based approaches showed lowest effectiveness (2-3\% success rate), while lexical transformations like synonym replacement proved moderate effectiveness (3-11\% success rate). For comparison, watermarked text had a mean g-score of 0.563, while unwatermarked text averaged 0.501.

\begin{table*}[h]
\centering

\label{tab:synthid_comprehensive_metrics}
\begin{tabularx}{\textwidth}{Xcccccccc}
\toprule
\textbf{Attack Method} & \multicolumn{2}{c}{\textbf{BERT Score}} & \multicolumn{2}{c}{\textbf{Complexity}} & \multicolumn{2}{c}{\textbf{Readability}} & \multicolumn{2}{c}{\textbf{Error}} \\
\cmidrule(lr){2-3}\cmidrule(lr){4-5}\cmidrule(lr){6-7}\cmidrule(lr){8-9}
 & \textbf{Avg} & \textbf{\% $\updownarrows$} & \textbf{Avg} & \textbf{\% $\updownarrows$} & \textbf{Avg} & \textbf{\% $\updownarrows$} & \textbf{Avg} & \textbf{\% $\updownarrows$} \\
\midrule
Watermarked (Ref.) & 1.000 & --- & 6.57 & --- & 79.71 & --- & 2.27 & --- \\
Unwatermarked & --- & --- & 7.37 & +12.18 & 75.58 & -5.18 & 2.68 & +18.06 \\
Transl. (German) & 0.939 & -6.12 & 6.80 & +3.50 & 80.03 & +0.40 & 2.17 & -4.41 \\
Transl. (Spanish) & 0.942 & -5.84 & 7.58 & +15.37 & 71.40 & -10.43 & 2.73 & +20.26 \\
Transl. (French) & 0.935 & -6.52 & 6.93 & +5.48 & 78.37 & -1.69 & 2.14 & -5.73 \\
Simple Paraphrase & 0.901 & -9.91 & 8.00 & +21.77 & 73.66 & -7.60 & 24.07 & +960.35 \\
BART Paraphrase & 0.900 & -10.0 & 6.69 & +1.83 & 80.15 & +0.54 & 0.90 & -60.35 \\
Pegasus Paraphrase & 0.872 & -12.8 & 6.08 & -7.46 & 81.86 & +2.70 & 0.06 & -97.36 \\
Swap & 0.899 & -10.0 & 6.52 & -0.76 & 79.71 & +0.00 & 9.93 & +337.44 \\
Deletion & 0.890 & -11.0 & 6.42 & -2.28 & 80.04 & +0.41 & 5.50 & +142.29 \\
Syn. Replacement & 0.894 & -10.5 & 8.39 & +27.70 & 72.28 & -9.32 & 24.39 & +974.45 \\
\bottomrule
\end{tabularx}
\caption{Comprehensive Text Characteristic Analysis by Attack Method for SynthID Watermarking}
\label{table:text_characteristic_synthid}
\end{table*}

Table~\ref{table:text_characteristic_synthid} presents detailed text quality metrics for each attack method applied to SynthID watermarked text. Translation-based attacks maintained reasonable semantic preservation (5.8-6.5\% BERT reduction) while modifying text structure enough to occasionally evade detection. Word swapping and deletion preserved textual metrics well but achieved limited watermark removal. Neural paraphrasing techniques again demonstrated the optimal balance, with Pegasus achieving 14\% removal success while limiting BERT reduction to 12.8\% and maintaining minimal grammatical errors (97.3\% reduction versus reference text).

Figures~\ref{fig:text_qualitye}, ~\ref{fig:text_qualityf}, ~\ref{fig:text_qualityg}, ~\ref{fig:text_qualityh} show the incremental effects of text modifications on SynthID Watermarking text quality metrics. SynthID demonstrates intermediate resistance to incremental attacks, with watermark degradation occurring more gradually than in Publicly Detectable Watermarking but more rapidly than in Provable Robust Watermarking. The sharp initial drops in BERT scores (second row) for all attack types indicate that even minor modifications (5-10\% of text) significantly impact semantic preservation while beginning to degrade watermark detectability. Synonym replacement shows particularly problematic quality trade-offs, with grammatical errors increasing exponentially after 15\% modification intensity while providing only moderate watermark removal benefits.

Figure~\ref{fig:watermark_analysis7} presents four violin plots which represent text quality metrics (BERT scores, complexity, readability, and grammatical errors), revealing significant text degradation from most attack methods. Figure~\ref{fig:watermark_analysis8} illustrate the incremental impact of increasing attack intensity on detection scores, demonstrating that even minimal modifications result in complete watermark removal for this scheme. Figure~\ref{fig:watermark_analysis9} shows the 100\% success rate for all attack strategies against this watermarking scheme, indicating complete vulnerability even to relatively simple text modifications. The visualisations collectively highlight that BART Paraphrase and Pegasus Paraphrase achieved the best balance between attack success and quality preservation, while maintaining relatively low grammatical error rates compared to lexical transformation approaches.

\begin{table}[ht]
\centering

\label{tab:pdw_effectiveness}
\begin{tabular}{lcc}
\hline
\textbf{Attack Method} & \textbf{\% Success Rate}\\
\hline
BART Paraphrase & 100.0 \\
Deletion & 100.0\\
Pegasus Paraphrase & 100.0 \\
Simple Paraphrase & 100.0 \\
Word Swap & 100.0\\
Synonym Replacement & 100.0\\
Translation (German) & 100.0  \\
Translation (Spanish) & 100.0 \\
Translation (French) & 100.0 \\
\hline
\end{tabular}
\caption{Comprehensive Analysis of Publicly Detectable Watermarking Attack Methods}
\label{table:publicly_detectable}
\end{table}

Table~\ref{table:publicly_detectable} demonstrates the complete vulnerability of the Publicly Detectable Watermarking scheme to all tested attack methods. Each attack strategy achieved a 100\% success rate in removing the watermark, indicating fundamental weaknesses in the cryptographic signature approach employed by this scheme.

\begin{table*}[ht]
\centering

\label{tab:pdw_quality}
\begin{tabularx}{\textwidth}{Xcccccccc}
\toprule
\textbf{Attack Method} & \multicolumn{2}{c}{\textbf{BERT Score}} & \multicolumn{2}{c}{\textbf{Complexity}} & \multicolumn{2}{c}{\textbf{Readability}} & \multicolumn{2}{c}{\textbf{Error}} \\
\cmidrule(lr){2-3}\cmidrule(lr){4-5}\cmidrule(lr){6-7}\cmidrule(lr){8-9}
 & \textbf{Avg} & \textbf{\% $\updownarrows$} & \textbf{Avg} & \textbf{\% $\updownarrows$}  & \textbf{Avg} & \textbf{\% $\updownarrows$} & \textbf{Avg} & \textbf{\% $\updownarrows$} \\
\midrule
Watermarked (Ref.) & 1.000 & --- & 10.67 & --- & 61.06 & --- & 20.45 & --- \\
Unwatermarked & --- & --- & 7.05 & -33.93 & 75.25 & +23.24 & 8.71 & -57.41 \\
Transl. (German) & 0.799 & -20.1 & 44.28 & +315.00 & -288.14 & -571.93 & 10.19 & -50.17 \\
Transl. (Spanish) & 0.827 & -17.3 & 24.25 & +127.27 & -22.64 & -137.08 & 5.86 & -71.34 \\
Transl. (French) & 0.833 & -16.7 & 23.16 & +117.06 & 11.06 & -81.89 & 5.68 & -72.22 \\
Simple Paraphrase & 0.824 & -17.6 & 12.74 & +19.40 & 46.44 & -23.93 & 104.34 & +410.22 \\
BART Paraphrase & 0.857 & -14.3 & 10.31 & -3.37 & 60.65 & -0.66 & 2.40 & -88.26 \\
Pegasus Paraphrase & 0.838 & -16.2 & 8.70 & -18.46 & 69.20 & +13.34 & 1.30 & -93.64 \\
Swap & 0.908 & -9.25 & 10.69 & +0.19 & 61.08 & +0.04 & 40.13 & +96.23 \\
Deletion & 0.875 & -12.5 & 10.72 & +0.47 & 61.58 & +0.86 & 24.25 & +18.58 \\
Syn. Replacement & 0.824 & -17.6 & 12.90 & +20.90 & 47.43 & -22.32 & 105.05 & +413.69 \\
\bottomrule
\end{tabularx}
\caption{Comprehensive Text Characteristic Analysis by Attack Method for Publicly Detectable Watermarking}
\label{table:text_characteristic_publicly_detectable}
\end{table*}

Table~\ref{table:text_characteristic_publicly_detectable} presents detailed text quality metrics for each attack method applied to Publicly Detectable watermarked text. Despite complete watermark removal success across all methods, the quality impact varied dramatically. Word swapping achieved watermark removal with minimal semantic degradation (9.25\% BERT reduction), while translation methods caused severe text distortion—German translation increased complexity by 315\% and decreased readability by 571.93\%. Neural paraphrasing methods (BART, Pegasus) offered favorable quality trade-offs with 14-16\% BERT reductions while removing watermarks with perfect success.

Figures~\ref{fig:text_qualityz}, ~\ref{fig:text_qualityy}, ~\ref{fig:text_qualityx}, ~\ref{fig:text_qualityw} shows the incremental effects of text modifications on Publicly Detectable Watermarking text quality metrics.  The relatively flat readability curves (third row) for swapping and deletion attacks suggest these methods preserve text fluency effectively despite successfully removing watermarks. In contrast, translation-based attacks (not shown in these incremental plots) caused severe readability degradation.

Our analysis reveals significant differences in the vulnerability of the three watermarking schemes. Provable Robust Watermarking demonstrated the strongest resistance to attacks, with only Pegasus Paraphrase achieving notable success (23\% removal rate) while other methods averaged just 3-5\% success. The mean Z-score for this scheme remained high (8.43) in watermarked text, with most attacks only moderately reducing it. SynthID showed moderate vulnerability, with Pegasus Paraphrase again proving most effective (14\% success rate), followed by synonym replacement (11\%) and word swaps (9\%). Most concerning was Publicly Detectable Watermarking, which demonstrated complete vulnerability with all attack methods achieving 100\% success in watermark removal.

Text quality impact varied significantly across attack methods. Pegasus Paraphrase consistently offered the best balance between watermark removal and quality preservation, with BERT score reductions of 12-16\% across schemes. Translation-based attacks maintained moderate semantic preservation (4-6\% BERT reduction for Provable Robust and SynthID) but occasionally produced severe readability issues, particularly with Publicly Detectable Watermarking where German translation increased text complexity by 315\%. Lexical transformations proved problematic for text quality, with synonym replacement causing grammatical error increases of 797\% (Provable Robust) and 974\% (SynthID). Simple paraphrase methods resulted in the highest quality degradation overall, with BERT score reductions of 16-17\% and substantial increases in grammatical errors.

The incremental effects analysis demonstrated a clear tradeoff between watermark removal and text quality. Progressive text modifications gradually reduced watermark detectability while simultaneously degrading semantic preservation (increased grammatical errors, decreased BERT scores, readability and complexity)in a near-linear relationship. This was particularly evident with SynthID, where even small modifications significantly reduced detection scores but incrementally diminished BERT scores. Notably, Publicly Detectable Watermarking showed complete vulnerability even to minimal modifications, while Provable Robust required substantial text alteration to bypass detection thresholds. These findings highlight that while current watermarking schemes can be defeated with varying degrees of effort, successful attacks generally require noticeable text modifications that compromise quality, suggesting that hybrid detection approaches incorporating stylometric and quality analysis could enhance watermark robustness.

\section{Discussion}
The results of this study uncover alarming gaps related to the current watermarking techniques based on large language models (LLMs), which will certainly impact content authentication technologies. The contrasts between Provable Robust Watermarking and the relatively weaker Publicly Detectable Watermarking accentuate the need to rethink some guiding design principles. The differing responses to attacks indicate that watermarking schemes prioritizing distributional modifications (Provable Robust) are more resilient to semantic-preserving attacks than those relying on structure-based signatures (Publicly Detectable).

The uniform success of neural paraphrasing techniques, especially Pegasus Paraphrase, on all schemes marks a significant gap in the effectiveness of watermarking methods. These sophisticated language models circumvent detection frameworks because they alter the expression of the underlying text while retaining its meaning. At the same time, approaches involving synonym substitution and other lexical shifts were shown to be less effective in removing watermarks while inflicting greater damage to the text, implying that watermarking techniques should focus on obstructing neural paraphrasing rather than straightforward manipulations of vocabulary.

Our analysis of text quality impacts reveals an important inverse relationship between attack effectiveness and output quality. Successful watermark removal invariably degraded semantic preservation, readability, or grammatical correctness. This trade-off suggests potential for hybrid detection systems that incorporate quality metrics alongside watermark detection, flagging content with suspicious combinations of marginal watermark scores and degraded quality indicators. Such approaches could significantly improve overall detection robustness without modifying the watermarking schemes themselves.

Despite claims of provable robustness, the watermarking schemes succumb to attacks due to the fundamental challenge of preserving semantic meaning while altering text structure. The effectiveness of attacks like neural paraphrasing, synonym substitution, and machine translation stems from their ability to reshape surface-level token distributions while maintaining coherence. Even probabilistic watermarking schemes, which modify token probabilities during generation, rely on statistical patterns that can be disrupted by paraphrasing. Cryptographic approaches, such as publicly detectable watermarking, fail when adversaries introduce minor perturbations that shift the encoded signature beyond detection limits. It is worth noting that the runtime of these attacks is relatively short as well—approximately 3 minutes per sample for Provable Robust Watermarking, and 6 minutes per sample for both Publicly Detectable Watermarking and SynthID Watermarking. This efficiency further increases the practical viability of such attacks. Ultimately, while watermarking aims to introduce imperceptible yet detectable modifications, adaptive transformations exploit the inherent flexibility of language, erasing these signals without compromising text utility.

Our study has some limitations. It was performed on a relatively small dataset of 100 text prompts, which may not capture the full scope of variability from a real-world LLM output. However, we made up for it by ensuring that the text is sufficiently long. In the future, it is important to focus on creating adaptive watermarking methods that detect and defend against paraphrasing attacks. Further work should investigate the possibility of embedding watermarks into deeper conceptual regions instead of superficial textual features. Furthermore, watermarking designs should allow adjustable variable strength settings to balance security and text quality needs for different applications. For immediate practical implementation, Provable Robust Watermarking offers the most secure option, though users should recognize that even this scheme remains vulnerable to sophisticated attacks. These findings underscore the ongoing challenge of maintaining trustworthy content attribution in increasingly adversarial environments.

\section{Conclusion}
The work presented in this paper indicates that recently developed watermarking techniques for LLMs are still at risk from semantic-preserving attacks of varying severity. According to our results, Provable Robust Watermarking appears to be the most resistant against attacks, whereas Publicly Detectable Watermarking is fully vulnerable to all attack methods. Most effective at watermark removal while sustaining the required text quality-neural paraphrasing posed a major threat to watermarking systems. As we analysed the watermarks, it became evident that there was an unavoidable trade-off between watermark removal efficiency and text fidelity, indicating possibilities for hybrid approaches that utilise quality measures in the detection process. This evidence necessitates the need for more advanced watermarking approaches that embed the signals within the deep semantic structures instead of surface-level text features. Technologies for watermarking in the future will need to be designed to withstand neural paraphrasing if they are to ensure reliable attribution of content against increasingly hostile attack vectors.

\section*{Acknowledgments}

\bibliography{custom}

\appendix

\section{Appendix}

\subsection{Provable Robust Watermarking Results}
\begin{figure}[H]
\centering
\includegraphics[width=\linewidth]{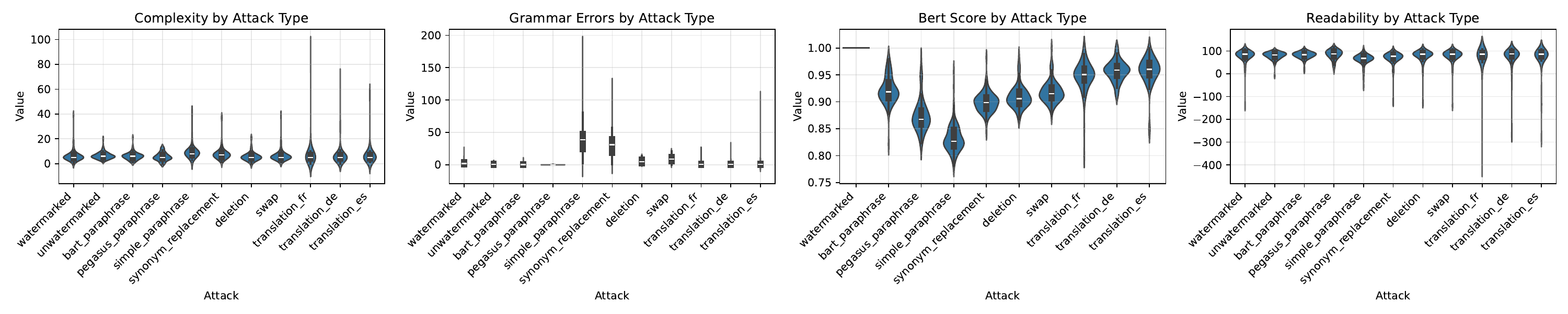}
\caption{Comprehensive results of Provable Robust Watermarking text quality scores across attack methods.}
\label{fig:watermark_analysisa}
\end{figure}

\begin{figure}[H]
\centering
\includegraphics[width=\linewidth]{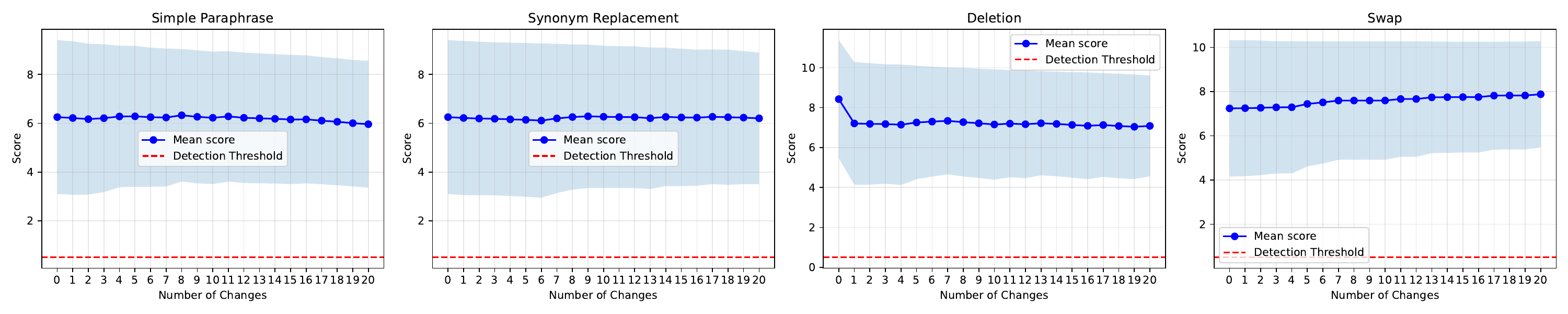}
\caption{Incremental effects of text modifications on detection rate for Provable Robust Watermarking across attack methods.}
\label{fig:watermark_analysisb}
\end{figure}

\begin{figure}[H]
\centering
\includegraphics[width=0.7\linewidth]{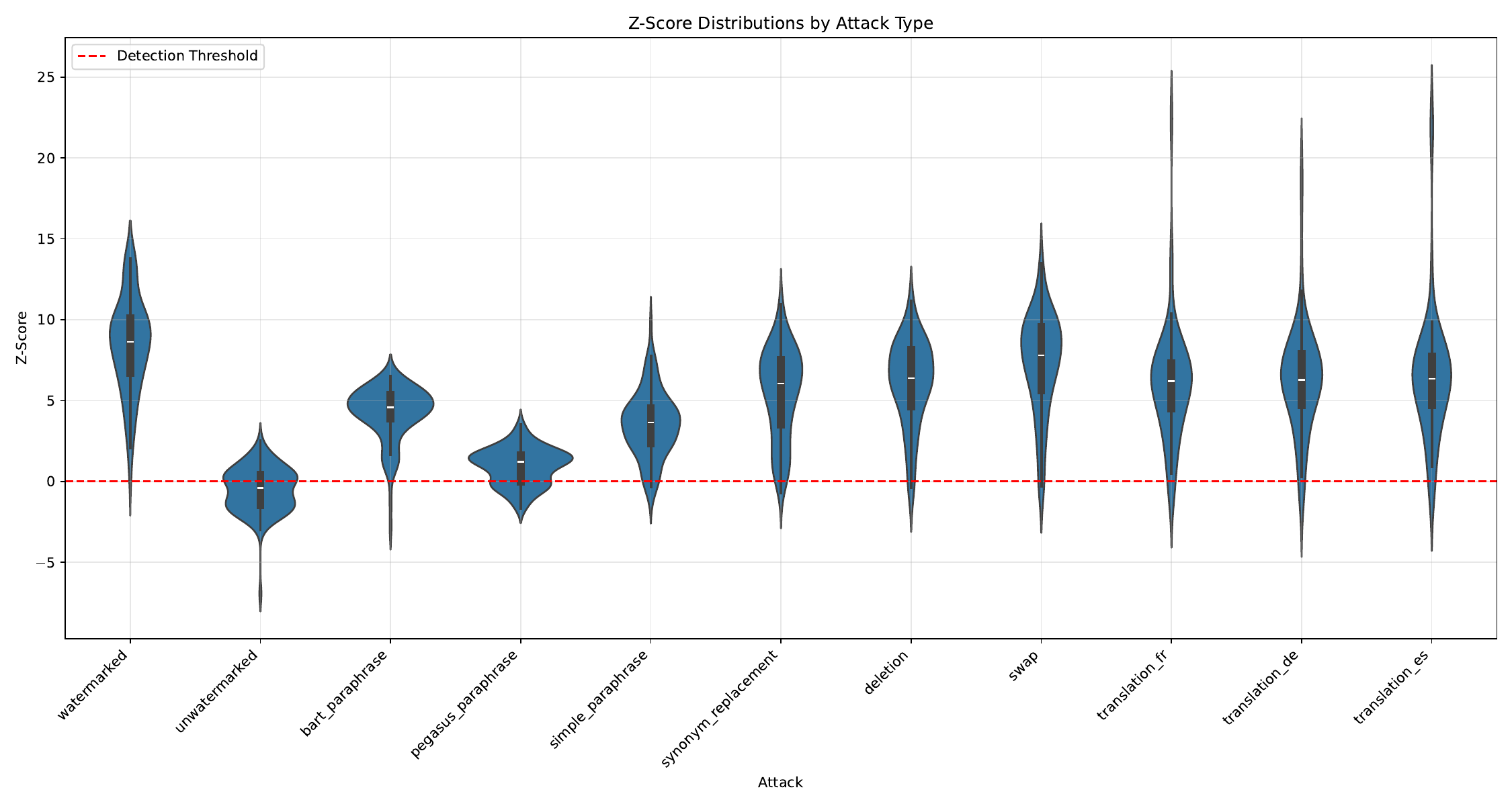}
\caption{Comprehensive results of Provable Robust Watermarking score distributions across attack methods.}
\label{fig:watermark_analysisc}
\end{figure}

\begin{figure}[H]
\centering
\includegraphics[width=\linewidth]{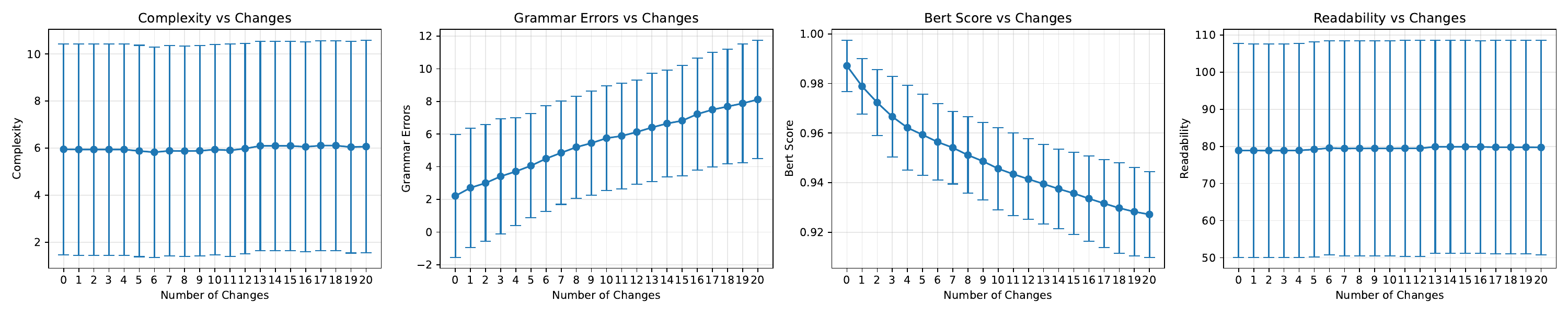}
\caption{Incremental effects of text modifications on text quality for Provable Robust Watermarking (Swap)}
\label{fig:text_qualitya}
\end{figure}

\begin{figure}[H]
\centering
\includegraphics[width=\linewidth]{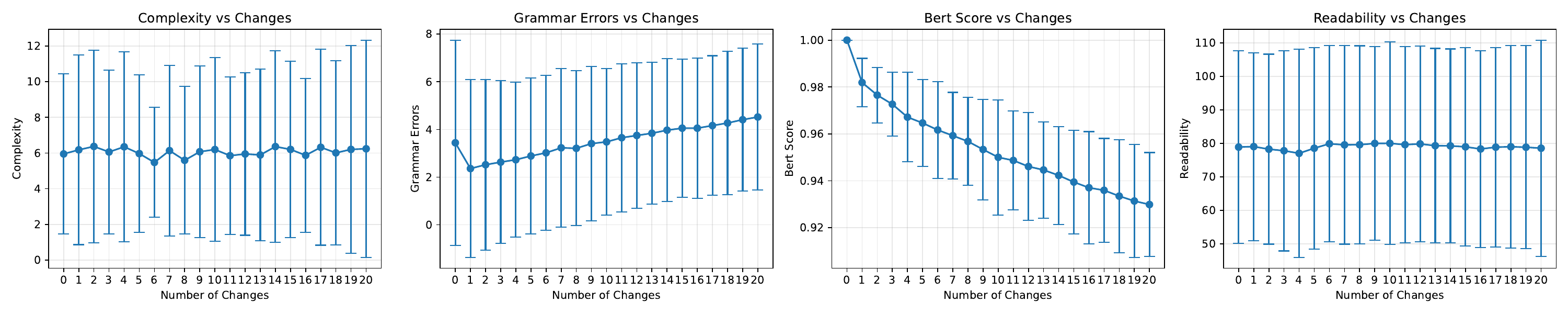}
\caption{Incremental effects of text modifications on text quality for Provable Robust Watermarking (Deletion)}
\label{fig:text_qualityb}
\end{figure}

\begin{figure}[H]
\centering
\includegraphics[width=\linewidth]{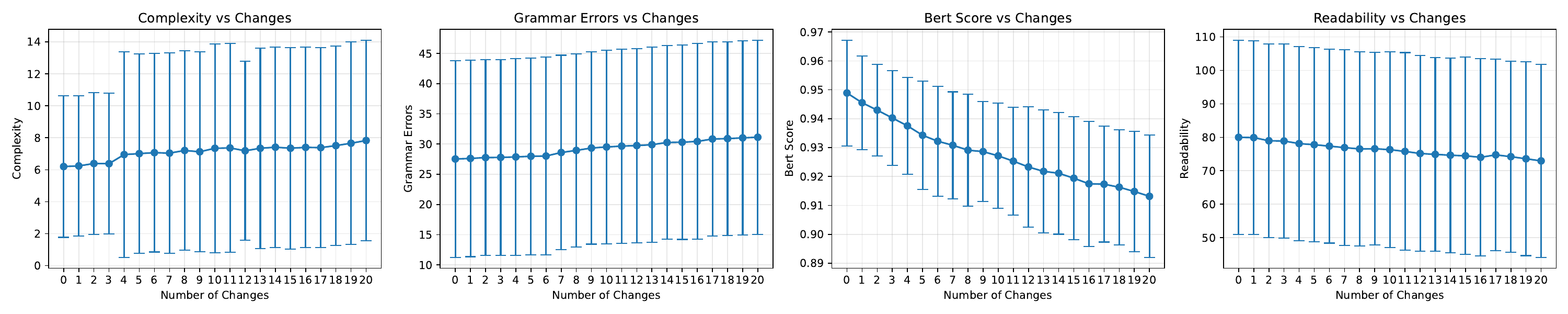}
\caption{Incremental effects of text modifications on text quality for Provable Robust Watermarking (Synonym Replacement)}
\label{fig:text_qualityc}
\end{figure}

\begin{figure}[H]
\centering
\includegraphics[width=\linewidth]{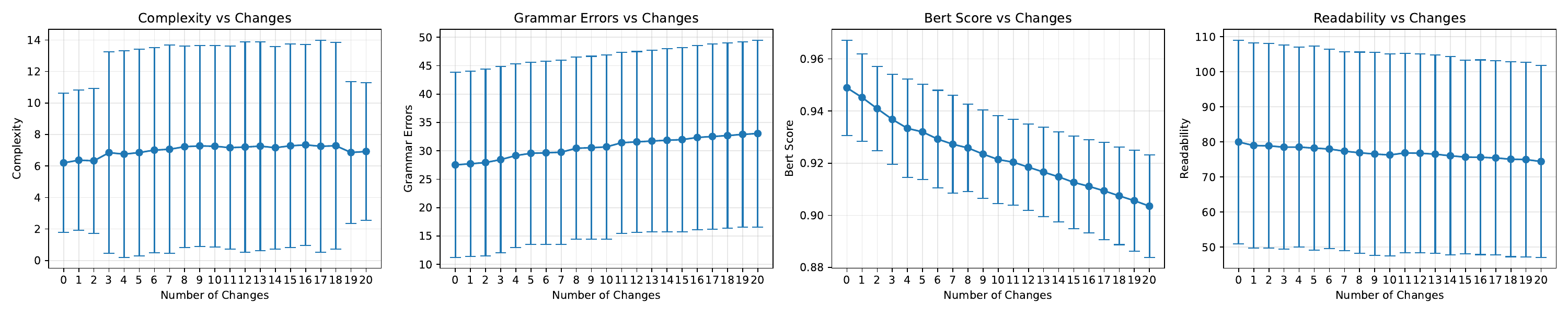}
\caption{Incremental effects of text modifications on text quality for Provable Robust Watermarking (Simple Replacement)}
\label{fig:text_qualityd}
\end{figure}

\subsection{SynthID-Text Results}
\begin{figure}[H]
\centering
\includegraphics[width=\linewidth]{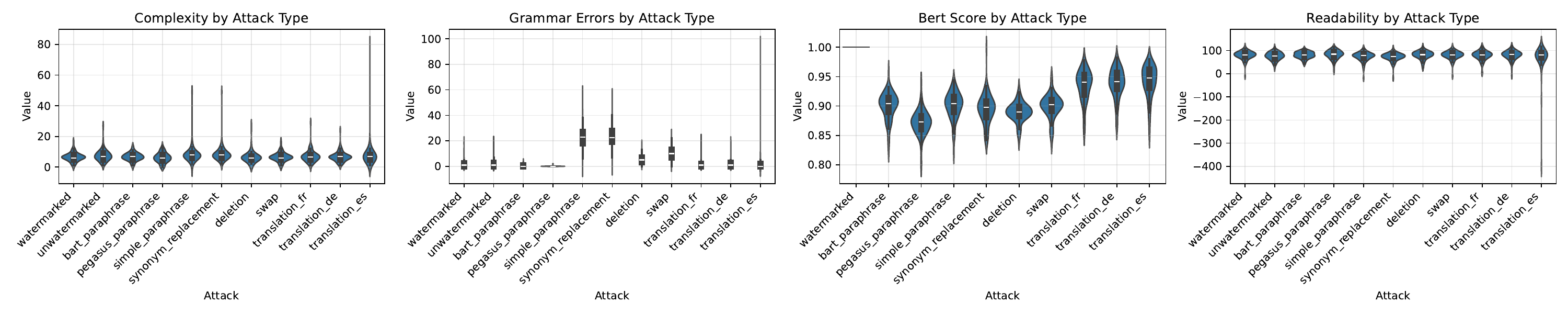}
\caption{Comprehensive results of SynthID Watermarking text quality scores across attack methods.}
\label{fig:watermark_analysis4}
\end{figure}

\begin{figure}[H]
\centering
\includegraphics[width=\linewidth]{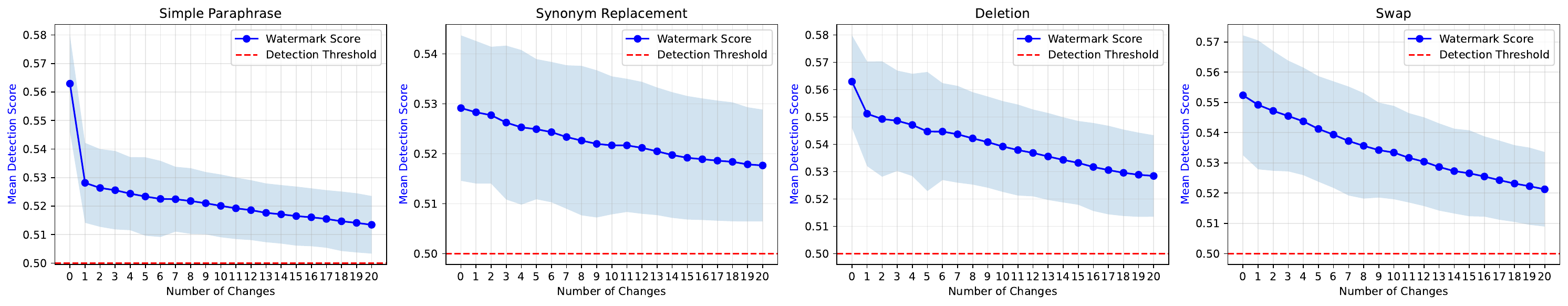}
\caption{Incremental effects of text modifications on detection rate for SynthID Watermarking across attack methods.}
\label{fig:watermark_analysis5}
\end{figure}

\begin{figure}[H]
\centering
\includegraphics[width=0.7\linewidth]{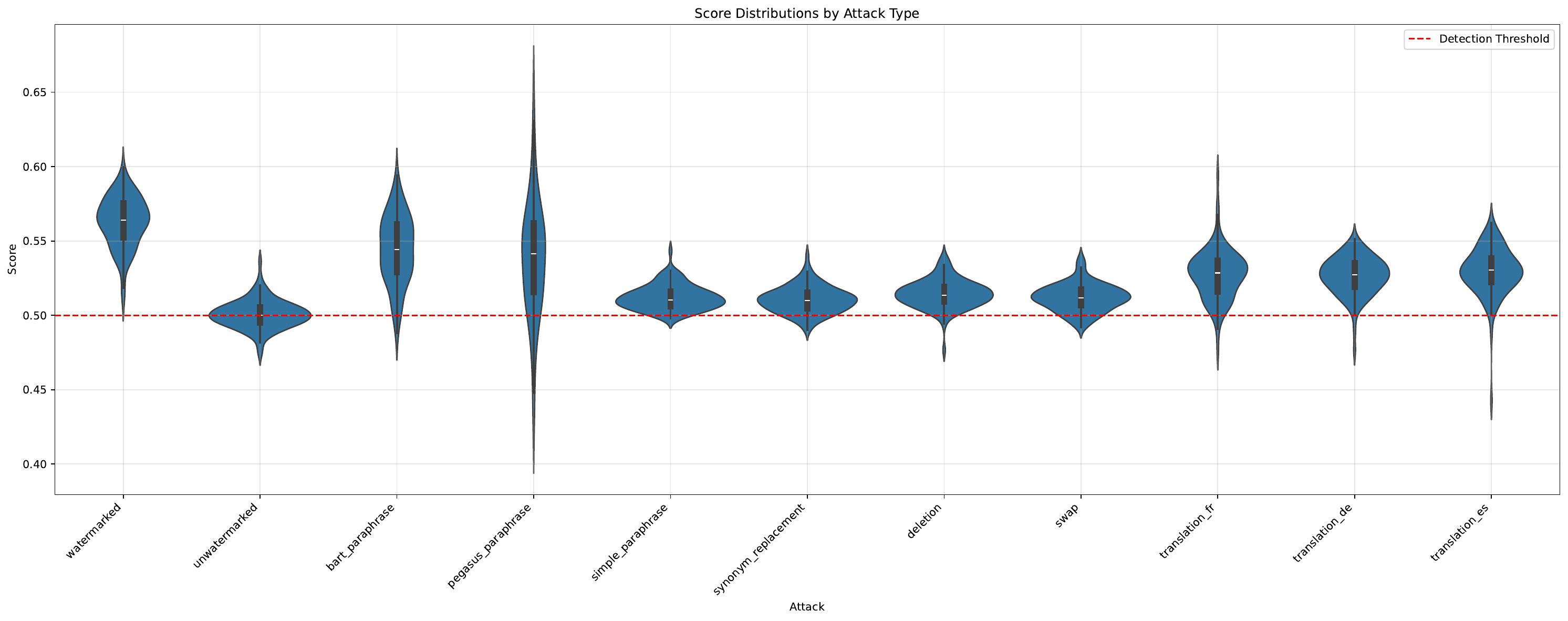}
\caption{Comprehensive results of SynthID Watermarking score distributions across attack methods.}
\label{fig:watermark_analysis6}
\end{figure}

\begin{figure}[H]
\centering
\includegraphics[width=\linewidth]{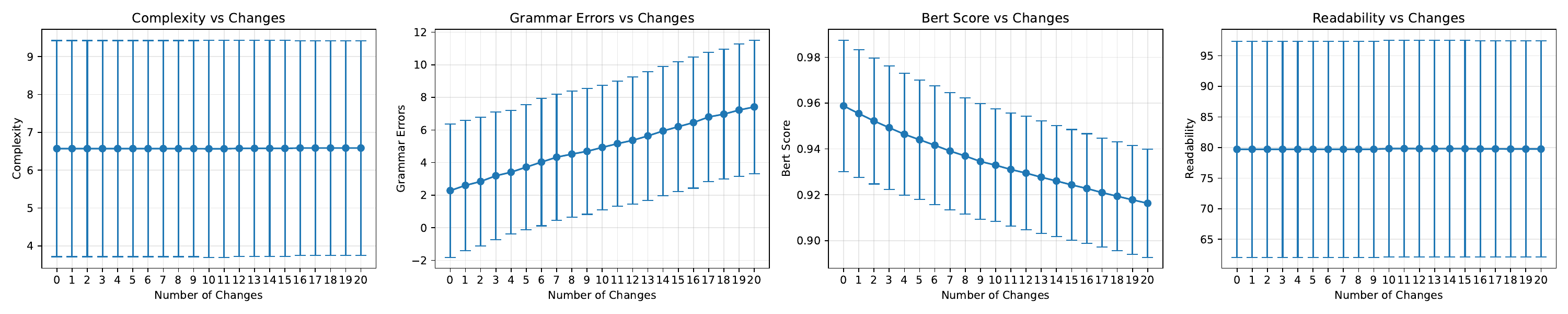}
\caption{Incremental effects of text modifications on text quality for SynthID Watermarking (Swap)}
\label{fig:text_qualitye}
\end{figure}

\begin{figure}[H]
\centering
\includegraphics[width=\linewidth]{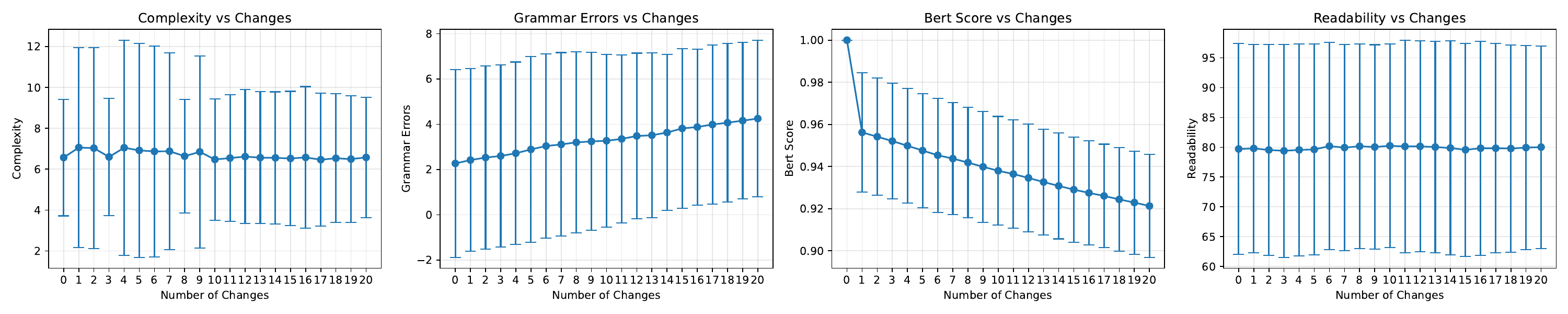}
\caption{Incremental effects of text modifications on text quality for SynthID Watermarking (Deletion)}
\label{fig:text_qualityf}
\end{figure}

\begin{figure}[H]
\centering
\includegraphics[width=\linewidth]{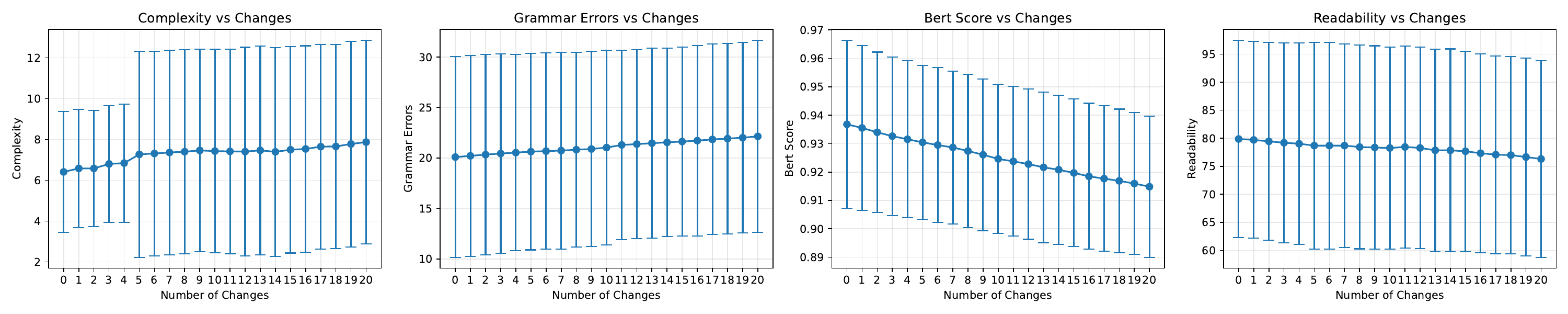}
\caption{Incremental effects of text modifications on text quality for SynthID Watermarking (Synonym Replacement)}
\label{fig:text_qualityg}
\end{figure}

\begin{figure}[H]
\centering
\includegraphics[width=\linewidth]{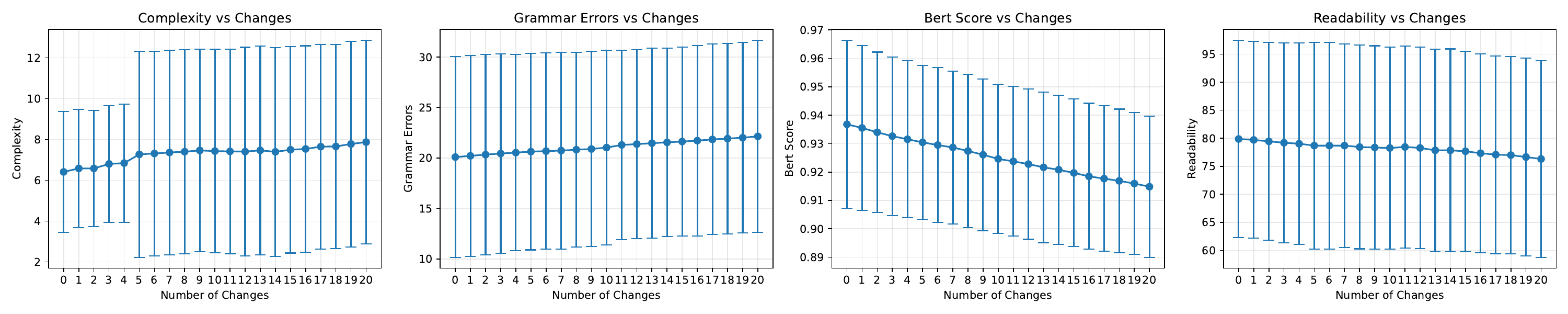}
\caption{Incremental effects of text modifications on text quality for SynthID Watermarking (Simple Replacement)}
\label{fig:text_qualityh}
\end{figure}

\subsection{Publicly Detectable Watermarking Results}

\begin{figure}[H]
\centering
\includegraphics[width=\linewidth]{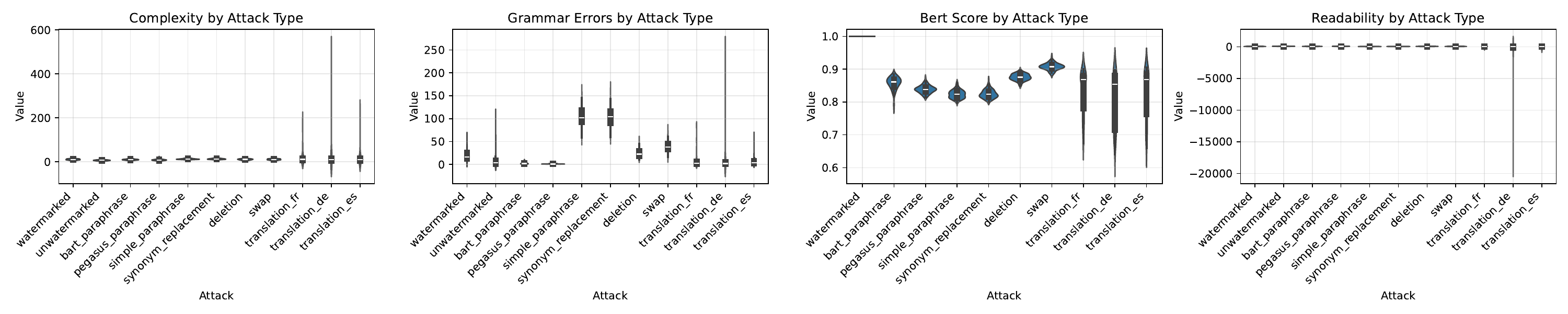}
\caption{Comprehensive results of Publicly Detectable Watermarking text quality scores across attack methods.}
\label{fig:watermark_analysis7}
\end{figure}

\begin{figure}[H]
\centering
\includegraphics[width=\linewidth]{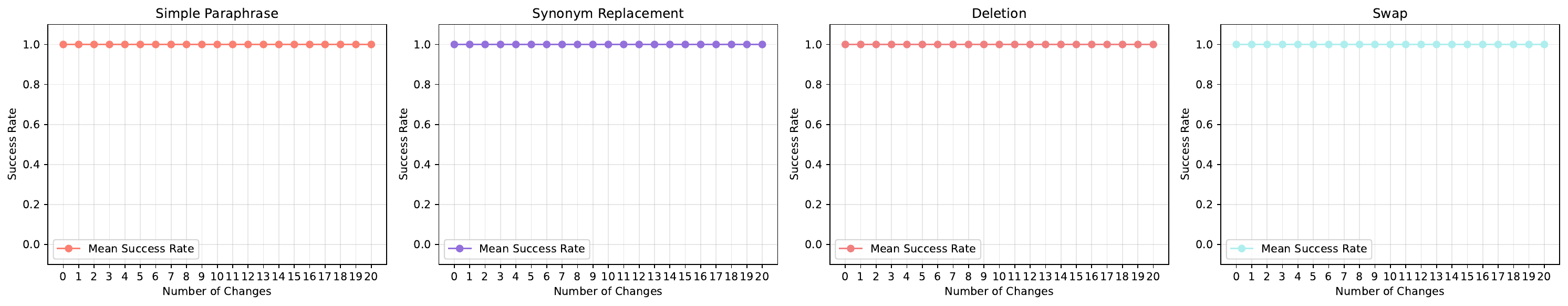}
\caption{Incremental effects of text modifications on detection rate for Publicly Detectable Watermarking across attack methods.}
\label{fig:watermark_analysis8}
\end{figure}

\begin{figure}[H]
\centering
\includegraphics[width=0.7\linewidth]{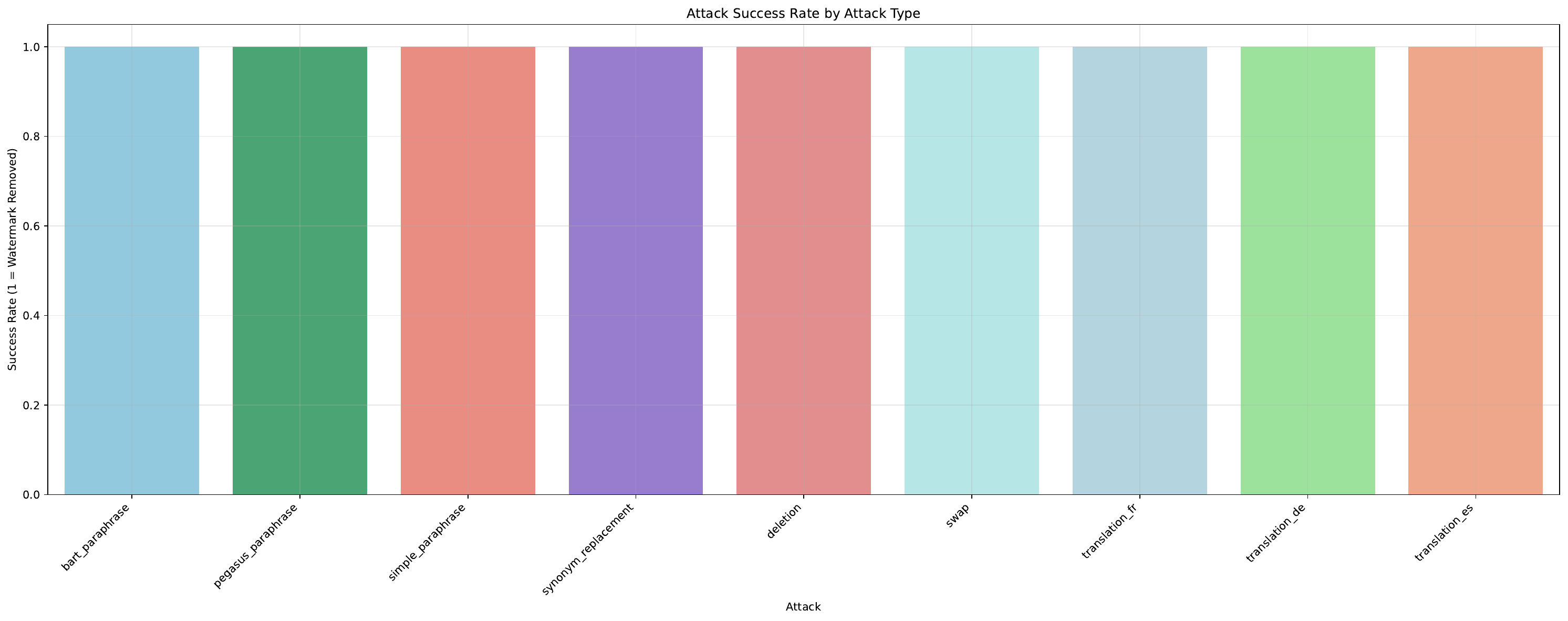}
\caption{Comprehensive results of Publicly Detectable Watermarking vulnerability across attack methods.}
\label{fig:watermark_analysis9}
\end{figure}

\begin{figure}[H]
\centering
\includegraphics[width=\linewidth]{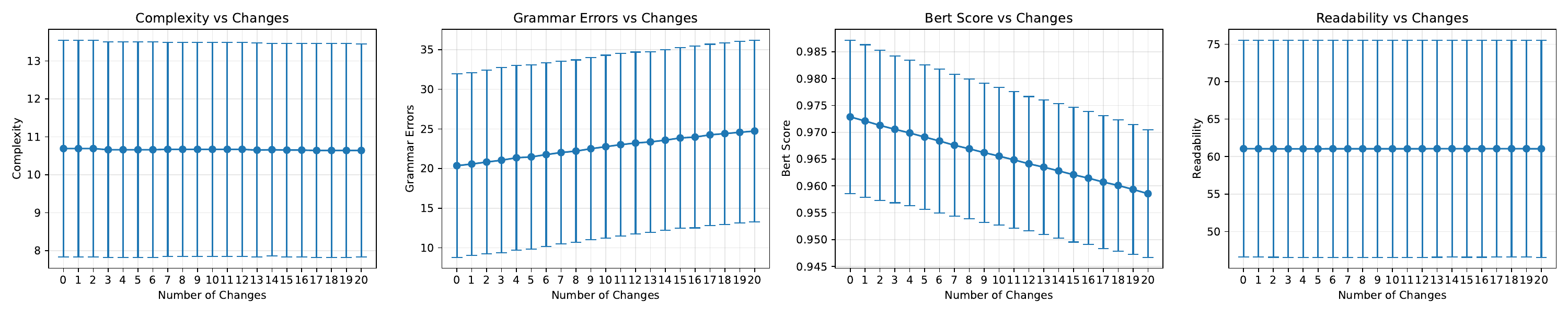}
\caption{Incremental effects of text modifications on text quality for Publicly Detectable Watermarking (Swap)}
\label{fig:text_qualityz}
\end{figure}

\begin{figure}[H]
\centering
\includegraphics[width=\linewidth]{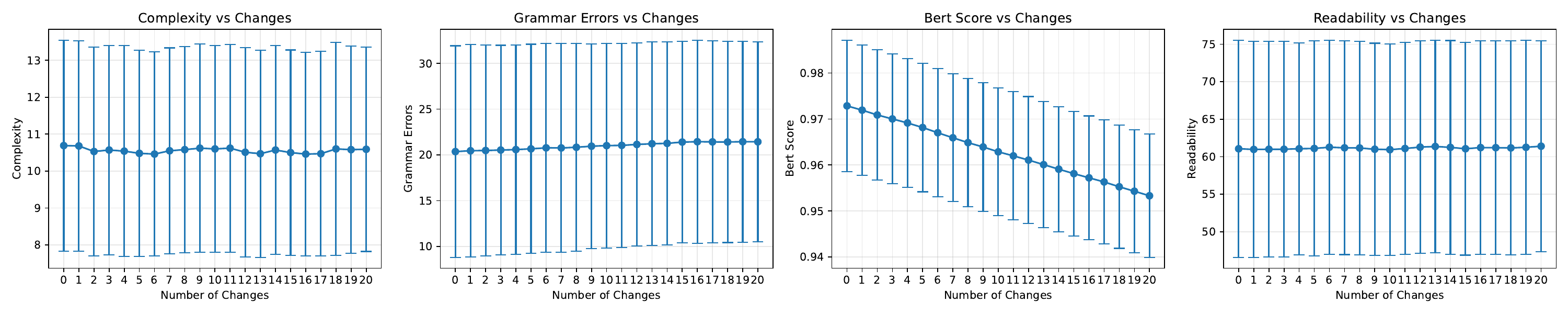}
\caption{Incremental effects of text modifications on text quality for Publicly Detectable Watermarking (Deletion)}
\label{fig:text_qualityy}
\end{figure}

\begin{figure}[H]
\centering
\includegraphics[width=\linewidth]{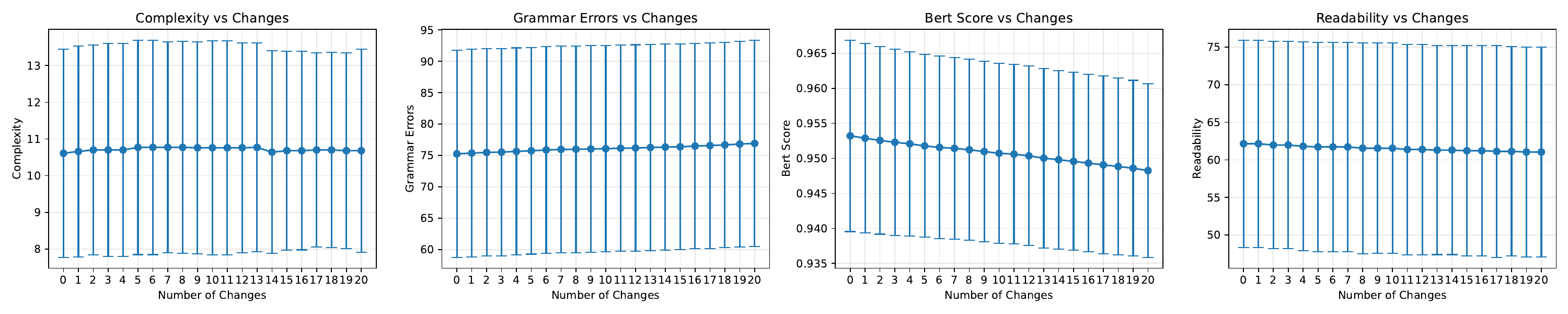}
\caption{Incremental effects of text modifications on text quality for Publicly Detectable Watermarking (Synonym Replacement)}
\label{fig:text_qualityx}
\end{figure}

\begin{figure}[H]
\centering
\includegraphics[width=\linewidth]{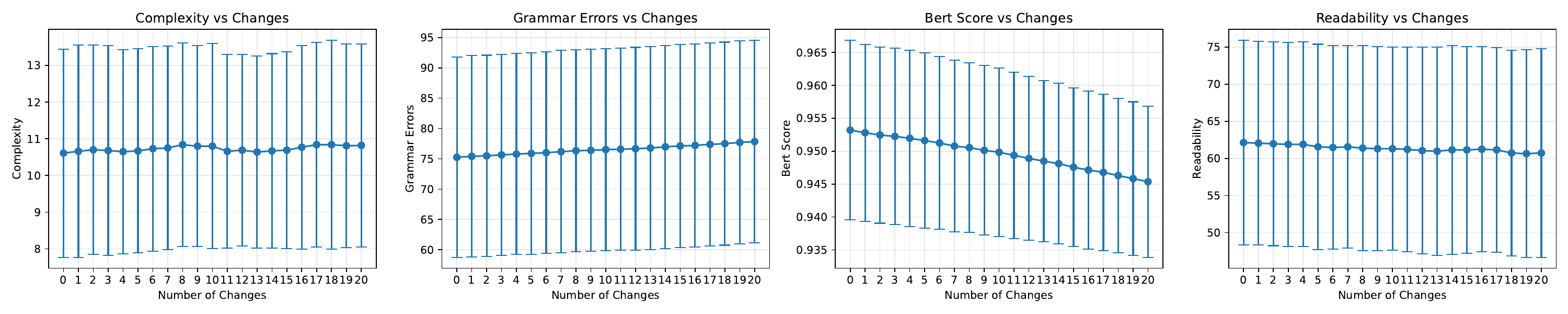}
\caption{Incremental effects of text modifications on text quality for Publicly Detectable Watermarking (Simple Replacement)}
\label{fig:text_qualityw}
\end{figure}

\end{document}